\documentclass[aps,pra,preprint,showpacs]{revtex4}
\newcommand{\PRA}[3]{Phys.~Rev.~A       {\bf #1}, #2 (#3).}

\begin{document}
\preprint{Version 1.0}

\title{Recursion relations for Hylleraas three-electron integral}

\author{Krzysztof Pachucki}
\email[]{krp@fuw.edu.pl}

\author{Mariusz Puchalski}
\email[]{mpuchals@fuw.edu.pl}

\affiliation{Institute of Theoretical Physics, Warsaw University,
             Ho\.{z}a 69, 00-681 Warsaw, Poland}

\author{Ettore Remiddi}
\email[]{remiddi@bo.infn.it}
\affiliation{Dipartimento di Fisica, Universit\`a di Bologna, 
and INFN Sezione di Bologna, I-40126 Bologna, Italy \\ 
CERN-PH-TH, CH-1211 Geneva 23, Switzerland }


\begin{abstract}
Recursion relations for Hylleraas three-electron integral are obtained in 
a closed form by using integration by parts identities. 
Numerically fast and well stable algorithm for the calculation of the 
integral with high powers of inter-electronic coordinates is presented. 
\end{abstract}

\maketitle
\section{Introduction}
The explicitely correlated Hylleraas basis set \cite{hyll} is one of the most efficient
representation of a few-electron wave function. It has been applied mostly to
helium \cite{hyl_hel} and lithium atoms \cite{hyl_lit}. 
While the general analytic formula for two-electron
Hylleraas integrals is quite simple \cite{kolos}, the three-electron case is much more
complicated. Fromm and Hill in \cite{fh} were able to derive a closed form formula
for the generic integral
\begin{equation}
\int d^3 r_1\,\int d^3 r_2\,\int d^3 r_3\,e^{-w_1\,r_1-w_2\,r_2-w_3\,r_3
-u_1\,r_{23}-u_2\,r_{31}-u_3\,r_{12}}\,
r_{23}^{-1}\,r_{31}^{-1}\,r_{12}^{-1}\,
r_{1}^{-1}\,r_{2}^{-1}\,r_{3}^{-1}\,, \label{01}
\end{equation}
which consists of the sum of many multi-valued dilogarithmic functions. This
formula was later simplified by Harris \cite{harris1} to a more symmetric form, in which many 
spurious singularities have been eliminated. However, only few
preliminary results have been reported so far with these functions, 
namely in Refs. \cite{rebane, harris2}.
The more convenient basis set for applications to 3- or 4-electron atoms 
happen to consist of functions with all $u_i$ equal to 0, and 
include polynomials in $r_{ij}$ and additionally in $r_i$, namely 
the 3-electron Hylleraas functions
\begin{equation}
\phi_{\{n\}} = e^{-w_1\,r_1-w_2\,r_2-w_3\,r_3}\,
r_{23}^{n_1-1}\,r_{31}^{n_2-1}\,r_{12}^{n_3-1}\,
r_{1}^{n_4-1}\,r_{2}^{n_5-1}\,r_{3}^{n_6-1}\,. \label{02}
\end{equation}
A compact expression for the integral with all $n_i = 0$ and with 
$n_1=2, n_2=0, n_3=2, n_4=n_5=n_6=1$ was derived in~\cite{remiddi}. 
No attempts have been reported so far to derive analytic formulas for
larger values of $n_i$. They  can in principle be obtained from the general
formula by Fromm and Hill in Ref. \cite{fh} by differentiation with respect 
to $u_i$ and $w_i$.
In practice that is quite complicated for large powers of $r_{ij}$. For testing
results obtained in this work we have written a Mathematica code for
derivation of Hylleraas type of integrals using the Fromm-Hill formula,
and obtained results for all $n_i$ such that $n_1+n_2+n_3+n_4+n_5+n_6<10$. 
Despite the existence of an analytical result for the generic integral 
(\ref{01}), the most advanced results for lithium
have been obtained so far by Yan and Drake in series of papers \cite{hyl_lit}, 
where Hylleraas integrals where calculated by series expansion in
angular momenta. 

For helium atom recursion relations \cite{kolos} are available, 
which allow to express Hylleraas integrals with $n_1+n_2+n_3=N$
[see Eq. (\ref{A5})]
as a linear combination of integrals from the lower shell $N-1$.
In this work we derive a similar complete set of recursion relations for 
the three-electron Hylleraas integrals
\begin{eqnarray}
f(n_1,n_2,n_3;n_4,n_5,n_6) &=& \int\frac{d^3\,r_1}{4\,\pi}\int\frac{d^3\,r_2}{4\,\pi}
\int\frac{d^3\,r_3}{4\,\pi}\,e^{-w_1\,r_1-w_2\,r_2-w_3\,r_3}\nonumber \\ &&
r_{23}^{n_1-1}\,r_{31}^{n_2-1}\,r_{12}^{n_3-1}\,
r_{1}^{n_4-1}\,r_{2}^{n_5-1}\,r_{3}^{n_6-1}\,. \label{03}
\end{eqnarray}
These recursions allow to  eliminate the principal drawback of Hylleraas basis
sets, namely the complexity of various integrals for nonrelativistic matrix elements
for the lithium atom. Moreover they can be applied 
also to more than three electron atoms with the condition that the basis 
set contains at most one odd power of $r_{ij}$. In the next section we present 
our main result, a complete set of recursion relations for the 
three-electron Hylleraas integrals defined in Eq. (\ref{03}).
In Section III  we analyze boundary terms and two-electron recursions,
and in Section IV we sketch the derivation of the main result.
Finally in Section V we analyze numerical stability and present 
numerical results for some selected integrals.  

\section{Three-electron recursion relations}
The recursion relations are conveniently divided into two sets. The first set 
of three recursions increases by 2 any of the first three indices 
$n_1,n_2,n_3$ at $n_4=n_5=n_6=0$, starting from the shell $n_1+n_2+n_3=0$ 
(as initial conditions, the values with $n_i=0,1, i=1,2,3$ 
and $n_4=n_5=n_6=0$ are therefore needed); 
with the notation of Appendix A, they read 
\begin{eqnarray}
f(n_1,n_2,n_3+2;0,0,0) &=&\frac{1+n_3}{2}\,\biggl\{
\nonumber \\ &&
\hspace*{-8mm}\frac{1}{w^2_1}\biggl[\frac{n_2\,(n_2-1)}{n_3+1}\,f(n_1,n_2-2,n_3+2;0,0,0)
\nonumber \\ &&
+(n_1+2\,n_2+n_3+2)\,f(n_1,n_2,n_3;0,0,0)
\nonumber \\ &&
+\frac{1}{n_3+1}\,f(n_1,n_2,n_3+2;\star,0,0)
\nonumber \\ &&
+\frac{n_2\,(n_2-1)}{n_1+1}\,f(n_1+2,n_2-2,n_3;0,0,0)
\nonumber \\ &&
+\frac{1}{n_1+1}\,f(n_1+2,n_2,n_3;0,0,\star)
\nonumber \\ &&
-\frac{\delta_{n_2}}{n_3+1}\,f(n_1,\star,n_3+2;0,0,0)
\nonumber \\ &&
-\frac{\delta_{n_2}}{n_1+1}\,f(n_1+2,\star,n_3;0,0,0)\biggr]
\nonumber \\ &&
\hspace*{-8mm}+\frac{1}{w^2_2}\biggl[\frac{n_1\,(n_1-1)}{n_3+1}\,f(n_1-2,n_2,n_3+2;0,0,0)
\nonumber \\ &&
+(2\,n_1+n_2+n_3+2)\,f(n_1,n_2,n_3;0,0,0)
\nonumber \\ &&
+\frac{1}{n_3+1}\,f(n_1,n_2,n_3+2;0,\star,0)+
\nonumber \\ &&
\frac{n_1\,(n_1-1)}{n_2+1}\,f(n_1-2,n_2+2,n_3;0,0,0)
+\nonumber \\ &&
\frac{1}{n_2+1}\,f(n_1,n_2+2,n_3;0,0,\star)
\nonumber \\ &&
-\frac{\delta_{n_1}}{n_3+1}\,f(\star,n_2,n_3+2;0,0,0)
\nonumber \\ &&
-\frac{\delta_{n_1}}{n_2+1}\,f(\star,n_2+2,n_3;0,0,0)\biggr]
\nonumber \\ &&
\hspace*{-13mm}-\frac{w^2_3}{w^2_1\,w^2_2}\biggl[\frac{n_3\,(n_3-1)}{n_2+1}\,f(n_1,n_2+2,n_3-2;0,0,0)
\nonumber \\ &&
+(n_1+n_2+2\,n_3+2)\,f(n_1,n_2,n_3;0,0,0)
\nonumber \\ &&
+\frac{1}{n_2+1}\,f(n_1,2+n_2,n_3;\star,0,0)
\nonumber \\ &&
+\frac{n_3\,(n_3-1)}{n_1+1}\,f(n_1+2,n_2,n_3-2;0,0,0)
\nonumber \\ &&
+\frac{1}{n_1+1}\,f(n_1+2,n_2,n_3;0,\star\,0)
\nonumber \\ &&
-\frac{\delta_{n_3}}{n_2+1}\,f(n_1,n_2+2,\star;0,0,0)
\nonumber \\ &&
-\frac{\delta_{n_3}}{n_1+1}\,f(n_1+2,n_2,\star;0,0,0)\biggr]\biggr\}\,.
\label{rec1}
\end{eqnarray}
Formulas for $f(n_1,n_2+2,n_3,0,0,0)$ and $f(n_1+2,n_2,n_3,0,0,0)$ can be
obtained from Eqs. (\ref{03}) and (\ref{rec1}) by interchanging $(r_2,r_3)$
or $(r_1,r_3)$ respectively. 

The next three recursions increase by 1 any of $n_4, n_5, n_6$ of a 
given subshell $n_4+n_5+n_6$ for arbitrary $n_1,n_2$ and $n_3$
\begin{eqnarray} 
f(n_1,n_2,n_3;n_4,n_5,n_6+1) &=&\frac{1}{w_1\,w_2\,w_3}\,\bigl\{
\nonumber \\ &&
(n_1-1)\,n_1\,n_4\,f(n_1-2,n_2,n_3;n_4-1,n_5+1,n_6)
\nonumber \\ &&
+(n_2-1)\,n_2\,n_5\,f(n_1,n_2-2,n_3;n_4+1,n_5-1,n_6)
\nonumber \\ &&
-(n_3-1)\,n_3\,n_4\,f(n_1,n_2,n_3-2;n_4-1,n_5+1,n_6)
\nonumber \\ &&
-(n_3-1)\,n_3\,n_5\,f(n_1,n_2,n_3-2;n_4+1,n_5-1,n_6)
\nonumber \\ &&
+n_4\,n_5\,(n_1+n_2-n_3+n_6+1)\,f(n_1,n_2,n_3;n_4-1,n_5-1,n_6)
\nonumber \\ &&
-(n_1-1)\,n_1\,\,f(n_1-2,n_2,n_3;n_4,n_5+1,n_6)\,w_1
\nonumber \\ &&
+(n_3-1)\,n_3\,\,f(n_1,n_2,n_3-2;n_4,n_5+1,n_6)\,w_1
\nonumber \\ &&
-n_5\,(n_1+n_2-n_3+n_6+1)\,f(n_1,n_2,n_3;n_4,n_5-1,n_6)\,w_1
\nonumber \\ &&
-(n_2-1)\,n_2\,f(n_1,n_2-2,n_3;n_4+1,n_5,n_6)\,w_2
\nonumber \\ &&
+(n_3-1)\,n_3\,f(n_1,n_2,n_3-2;n_4+1,n_5,n_6)\,w_2
\nonumber \\ &&
-n_4\,(n_1+n_2-n_3+n_6+1)\,f(n_1,n_2,n_3;n_4-1,n_5,n_6)\,w_2
\nonumber \\ &&
-n_4\,n_5\,f(n_1,n_2,n_3;n_4-1,n_5-1,n_6+1)\,w_3
\nonumber \\ &&
+(n_1+n_2-n_3+n_6+1)\,f(n_1,n_2,n_3;n_4,n_5,n_6)\,w_1\,w_2
\nonumber \\ &&
+n_4\,f(n_1,n_2,n_3;n_4-1,n_5,n_6+1)\,w_2\,w_3
\nonumber \\ &&
+n_5\,f(n_1,n_2,n_3;n_4,n_5-1,n_6+1)\,w_3\,w_1
\nonumber \\ &&
-\delta_{n_1}\,n_4\,f(\star,n_2,n_3;n_4-1,n_5+1,n_6)
\nonumber \\ &&
+\delta_{n_1}\,f(\star,n_2,n_3;n_4,n_5+1,n_6)\,w_1
\nonumber \\ &&
-\delta_{n_2}\,n_5\,f(n_1,\star,n_3;n_4+1,n_5-1,n_6)
\nonumber \\ &&
+\delta_{n_2}\,f(n_1,\star,n_3;n_4+1,n_5,n_6)\,w_2
\nonumber \\ &&
+\delta_{n_3}\,n_4\,f(n_1,n_2,\star;n_4-1,n_5+1,n_6)
\nonumber \\ &&
+\delta_{n_3}\,n_5\,f(n_1,n_2,\star;n_4+1,n_5-1,n_6)
\nonumber \\ &&
-\delta_{n_3}\,f(n_1,n_2,\star;n_4,n_5+1,n_6)\,w_1
\nonumber \\ &&
-\delta_{n_3}\,f(n_1,n_2,\star;n_4+1,n_5,n_6)\,w_2\bigr\}\,.
\label{rec2}
\end{eqnarray}
Two other recursions for $f(n_1,n_2,n_3,n_4,n_5+1,n_6)$ and $f(n_1,n_2,n_3,n_4+1,n_5,n_6)$
are obtained from Eqs. (\ref{03}) and (\ref{rec2}) by interchanging
arguments of $f$-function, using the identities
\begin{eqnarray}
&&f(n_1,n_2,n_3;n_4,n_5,n_6;w_1,w_2,w_3) = \nonumber \\
&&f(n_2,n_1,n_3;n_5,n_4,n_6;w_2,w_1,w_3) = \nonumber \\
&&f(n_3,n_2,n_1;n_6,n_5,n_4;w_3,w_2,w_1) = \nonumber \\
&&f(n_1,n_3,n_2;n_4,n_6,n_5;w_1,w_3,w_2)\,, \label{06} 
\end{eqnarray}
where any of $n_i$ can become a $\star$.

\section{Boundary terms}
These recursion relations in Eqs. (\ref{rec1}) and (\ref{rec2}) involve
boundary terms.  The first boundary term was derived in Ref. \cite{remiddi}, 
the 7 remaining boundary terms to start the first set of recursions
are obtained from the generic Fromm-Hill formula \cite{fh} or by direct 
integration
\begin{eqnarray}
f(0,0,0;0,0,0) &=& \frac{-1}{2\,w_1\,w_2\,w_3}\biggl\{\nonumber \\
&& \hspace*{-2ex}
\ln\biggl(\frac{w_3}{w_1+w_2}\biggr)\,\ln\biggl(1+\frac{w_3}{w_1+w_2}\biggr)
+{\rm Li}_2\biggl(-\frac{w_3}{w_1+w_2}\biggr)+{\rm Li}_2\biggl(1-\frac{w_3}{w_1+w_2}\biggr)
\nonumber \\
&& \hspace*{-4ex}
+\ln\biggl(\frac{w_2}{w_3+w_1}\biggr)\,\ln\biggl(1+\frac{w_2}{w_1+w_3}\biggr)
+{\rm Li}_2\biggl(-\frac{w_2}{w_1+w_3}\biggr)+{\rm Li}_2\biggl(1-\frac{w_2}{w_1+w_3}\biggr)
\nonumber \\
&& \hspace*{-4ex} 
+\ln\biggl(\frac{w_1}{w_2+w_3}\biggr)\,\ln\biggl(1+\frac{w_1}{w_1+w_2}\biggr)
+{\rm Li}_2\biggl(-\frac{w_1}{w_2+w_3}\biggr)+{\rm Li}_2\biggl(1-\frac{w_1}{w_2+w_3}\biggr)
\biggr\}\,,\nonumber \\
f(1,0,0;0,0,0) &=& -\frac{1}{w_2^2\,w_3^2}\,\ln\biggl[
                    \frac{w_1\,(w_1+w_2+w_3)}{(w_1+w_2)(w_1+w_3)}\biggr]\,,\nonumber \\
f(0,1,0;0,0,0) &=& -\frac{1}{w_1^2\,w_3^2}\,\ln\biggl[
                    \frac{w_2\,(w_1+w_2+w_3)}{(w_2+w_3)(w_2+w_1)}\biggr]\,,\nonumber \\
f(0,0,1;0,0,0) &=& -\frac{1}{w_1^2\,w_2^2}\,\ln\biggl[
                    \frac{w_3\,(w_1+w_2+w_3)}{(w_3+w_1)(w_3+w_2)}\biggr]\,,\nonumber \\
f(1,1,0;0,0,0) &=& \frac{1}{w_1\,w_2\,(w_1+w_2)\,w_3^2}\,,\nonumber \\
f(1,0,1;0,0,0) &=& \frac{1}{w_1\,w_3\,(w_1+w_3)\,w_2^2}\,,\nonumber \\
f(0,1,1;0,0,0) &=& \frac{1}{w_2\,w_3\,(w_2+w_3)\,w_1^2}\,,\nonumber \\
f(1,1,1;0,0,0) &=& \frac{1}{w_1^2\,w_2^2\,w_3^2}\,,
\label{3bound}
\end{eqnarray}
where Li$_2(x)$ is the dilogarithmic function.
Besides the initial terms one needs also as boundary terms the following 
two-electron integrals 
\begin{eqnarray}
f(\star,n_2,n_3;n_4,n_5,n_6) &=& \Gamma(n_5+n_6-1,n_4,n_3+n_2-1;w_2+w_3,w_1,0)\,,\nonumber \\
f(n_1,\star,n_3;n_4,n_5,n_6) &=& \Gamma(n_4+n_6-1,n_5,n_1+n_3-1;w_1+w_3,w_2,0)\,,\nonumber \\
f(n_1,n_2,\star;n_4,n_5,n_6) &=& \Gamma(n_4+n_5-1,n_6,n_1+n_2-1;w_1+w_2,w_3,0)\,,\nonumber \\
f(n_1,n_2,n_3;\star,0,0) &=& \Gamma(n_3-1,n_2-1,n_1;w_2,w_3,0)\,,\nonumber \\
f(n_1,n_2,n_3;0,\star,0) &=& \Gamma(n_1-1,n_3-1,n_2;w_3,w_1,0)\,,\nonumber \\
f(n_1,n_2,n_3;0,0,\star) &=& \Gamma(n_2-1,n_1-1,n_3;w_1,w_2,0)\,,
\label{2bound}
\end{eqnarray}
where $\Gamma$ is defined in Eq. ({\ref{A5}). Recursion relations for $\Gamma$
have already been worked out in the literature and can be found for example 
in \cite{kolos, kor1}. For completeness we include below
generic formulas for two cases: nonnegative $n_1,n_2,n_3$,
\begin{equation}
\Gamma(n_1,n_2,n_3;\alpha_1,\alpha_2,\alpha_3) = 
\biggl(-\frac{\rm d}{{\rm d}\alpha_1}\biggr)^{n_1}\,
\biggl(-\frac{\rm d}{{\rm d}\alpha_2}\biggr)^{n_2}\,
\biggl(-\frac{\rm d}{{\rm d}\alpha_3}\biggr)^{n_3}\,
\frac{1}{(\alpha_1+\alpha_2)(\alpha_2+\alpha_3)(\alpha_3+\alpha_1)},
\label{09}
\end{equation}
and the second case: $n_1=-1$, and nonnegative $n_2,n_3$
\begin{equation} 
\Gamma(-1,n_2,n_3;\alpha_1,\alpha_2,\alpha_3) = 
\biggl(-\frac{\rm d}{{\rm d}\alpha_2}\biggr)^{n_2}\,
\biggl(-\frac{\rm d}{{\rm d}\alpha_3}\biggr)^{n_3}\,
\frac{\ln(\alpha_1+\alpha_2)-\ln(\alpha_1+\alpha_3)}
{(\alpha_2-\alpha_3)(\alpha_2+\alpha_3)}\,.
\label{10}
\end{equation}
The complete set of  recursion relations have been checked by comparison
with analytic expression obtained by differentiation of
Fromm and Hill formula \cite{fh} with respect to $u_i$ and $w_i$ 
parameters (in our notation).

\section{Derivation of three-electron recursion formulas}
We use the method of the integration by parts identities~\cite{fdiag}, 
which is by now standard in the analytical calculation of 
Feynman diagrams. In our case, it amounts to consider the following 9 
identities in the momentum representation of the integral $G$, 
[see Eq. (\ref{A6})]
\begin{eqnarray}
&&0 \equiv {\rm id}(i,j) = 
\int d^3k_1\int d^3k_2\int d^3k_3\,\frac{\partial}{\partial\,{\vec k_i}}
 \Bigl[ \vec k_j\,(k_1^2+u_1^2)^{-m_1} 
\nonumber \\ &&  
(k_2^2+u_2^2)^{-m_2}\,(k_3^2+u_3^2)^{-m_3}
(k_{32}^2+w_1^2)^{-m_4}\,(k_{13}^2+w_2^2)^{-m_5}\,(k_{21}^2+w_3^2)^{-m_6} 
\Bigr] , 
\label{id}
\end{eqnarray} 
which are trivially valid (the integral of the derivative of a function 
vanishing at infinity vanishes). \par 
These identities group naturally into three sets. The first set consists of 
id$(1,1)$, id$(2,1)$, and id$(3,1)$. Other sets are obtained by changing
the second argument from $1$ into $2$ or $3$.
The reduction of the scalar products from numerator leads to the following
identities of the first set
\begin{eqnarray}
{\rm id}(1,1)&=& 
-m_6\,G(m_1-1,m_6+1) - m_5\,G(m_1-1,m_5+1) + m_6\,G(m_2-1,m_6+1) 
\nonumber \\ &&
+ m_5\,G(m_3-1,m_5+1) + (3-2\,m_1-m_5-m_6)\,G() + m_6\,G(m_6+1)\,u^2_1 
\nonumber \\ &&
+ m_5\,G(m_5+1)\,u^2_1 + 2\,m_1\,G(m_1+1)\,u^2_1 - m_6\,G(m_6+1)\,u^2_2 - m_5\,G(m_5+1)\,u^2_3 
\nonumber \\ &&
+ m_5\,G(m_5+1)\,w^2_2 + m_6\,G(m_6+1)\,w^2_3\,,
\nonumber \\ \nonumber \\ 
{\rm id}(2,1)&=& 
-m_6\,G(m_1-1,m_6+1) - m_5\,G(m_1-1,m_5+1) + m_6\,G(m_2-1,m_6+1) 
\nonumber \\ &&
+ m_5\,G(m_3-1,m_5+1) - m_5\,G(m_4-1,m_5+1) + (m_6-m_1)\,G() 
\nonumber \\ &&
+ m_5\,G(m_5+1, m_6-1) - m_1\,G(m_1+1, m_2-1) + m_1\,G(m_1+1,m_6-1) 
\nonumber \\ &&
+ m_6\,G(m_6+1)\,u^2_1 + m_5\,G(m_5+1)\,u^2_1 + m_1\,G(m_1+1)\,u^2_1 - m_6\,G(m_6+1)\,u^2_2 
\nonumber \\ &&
+ m_1\,G(m_1+1)\,u^2_2 - m_5\,G(m_5+1)\,u^2_3 + m_5\,G(m_5+1)\,w^2_1 - m_6\,G(m_6+1)\,w^2_3 
\nonumber \\ &&
- m_5\,G(m_5+1)\,w^2_3 - m_1\,G(m_1+1)\,w^2_3\,,
\nonumber \\ \nonumber \\ 
{\rm id}(3,1)&=& 
-m_6\,G(m_1-1,m_6+1) - m_5\,G(m_1-1,m_5+1) + m_6\,G(m_2-1,m_6+1) 
\nonumber \\ &&
+ m_5\,G(m_3-1,m_5+1) - m_6\,G(m_4-1,m_6+1) + m_6\,G(-m_5+1, m_6+1) 
\nonumber \\ &&
+ (m_5-m_1)\,G() - m_1\,G(m_1+1,m_3-1) + m_1\,G(m_1+1,m_5-1) 
\nonumber \\ &&
+ m_6\,G(m_6+1)\,u^2_1  + m_5\,G(m_5+1)\,u^2_1 + m_1\,G(m_1+1)\,u^2_1 -
m_6\,G(m_6+1)\,u^2_2 
\nonumber \\ &&
- m_5\,G(m_5+1)\,u^2_3 + m_1\,G(m_1+1)\,u^2_3 + m_6\,G(m_6+1)\,w^2_1 -
m_6\,G(m_6+1)\,w^2_2 
\nonumber \\ &&
- m_5\,G(m_5+1)\,w^2_2 - m_1\,G(m_1+1)\,w^2_2\,.
\label{ids}
\end{eqnarray}
The function $G$ is defined as in Eq. (\ref{A6}), but for ease of 
writing only the arguments which contain a $\pm 1$ are shown explicitly 
[so that, for instance, $G(m_5+1)$ stands for 
$G(m_1,m_2,m_3;m_4,m_5+1,m_6)$ etc.]
\par 
The general solution of these recursions is on itself of great interest,
but we consider here only the case
$u_1=u_2=u_3=0$ which corresponds to Hylleraas basis set.
For this we put $m_1=m_2=m_3=1$ and differentiate these identities
over $u_1,u_2,u_3$ at $u_1=u_2=u_3=0$. It leads to recursions for 
the $h$ function of 
Eq. (\ref{A7}). The first subset of identities forms 3 linear equations for
$h(n_1+2), h(m_5+1), h(m_6+1)$ (same convention for the explicitly written 
indices as for the function $G$) 
which can easily be solved. Since there are 9 equations
for 6 unknowns, two different solutions for $h(m_4+1), h(m_5+1), h(m_6+1)$ 
can be used to simplify 
recursion formulas. We achieve this by solving three equations
\begin{eqnarray}
h_1(m_4+1)-h_2(m_4+1) &=& 0\,,\nonumber \\
h_1(m_5+1)-h_2(m_5+1) &=& 0\,,\nonumber \\
h_1(m_6+1)-h_2(m_6+1) &=& 0\,,
\label{heqs}
\end{eqnarray}
against the following linear combinations
\begin{eqnarray}
h(m_5-1,m_6+1) - h(m_4-1,m_6+1)\,, && \nonumber \\
h(m_5+1,m_6-1) - h(m_4-1,m_5+1)\,, && \nonumber \\ 
h(m_4+1,m_5-1) - h(m_4+1,m_6-1)\,. &&
\label{hcomb}
\end{eqnarray}
The obtained solutions are inserted back in recursion formulas for $h(n_i+2)$ and $h(m_i+1)$,
and they take the form
\begin{eqnarray}
\frac{2\,w^2_2\,w^2_3}{n_1+1}\,h(n_1+2) &=& \bigl[w^2_3\,(2 + n_1 + n_2 + 2\,n_3)+
                                        w^2_2\,(2 + n_1 + 2\,n_2 + n_3)-\nonumber \\&&
                                        w^2_1\,(2 + 2\,n_1 + n_2 + n_3)\bigr]\,h()+w^2_3\,Q_3+w^2_2\,Q_2-w^2_1\,Q_1
                                        +\ldots \nonumber \\
\frac{2\,w^2_1\,w^2_3}{n_2+1}\,h(n_2+2) &=& \bigl[w^2_3\,(2 + n_1 + n_2 + 2\,n_3)-
                                        w^2_2\,(2 + n_1 + 2\,n_2 + n_3)+\nonumber \\&&
                                        w^2_1\,(2 + 2\,n_1 + n_2 + n_3)\bigr]\,h()+w^2_3\,Q_3-w^2_2\,Q_2+w^2_1\,Q_1
                                        +\ldots \nonumber \\
\frac{2\,w^2_1\,w^2_2}{n_3+1}\,h(n_3+2) &=& \bigl[-w^2_3\,(2 + n_1 + n_2 + 2\,n_3)+
                                        w^2_2\,(2 + n_1 + 2\,n_2 + n_3)+\nonumber \\&&
                                        w^2_1\,(2 + 2\,n_1 + n_2 + n_3)\bigr]\,h()-w^2_3\,Q_3+w^2_2\,Q_2+w^2_1\,Q_1
                                        +\ldots 
\label{h1solv}
\end{eqnarray}
\begin{eqnarray}
w^2_1\,m_4\,h(m_4+1)               &=& (- n_1 + n_2 + n_3 + 2\,m_4-1 )\,h()+2\,(X_3-X_2)+\ldots \nonumber \\
w^2_2\,m_5\,h(m_5+1)               &=& (n_1 - n_2 + n_3 + 2\,m_5-1 )\,h()+2\,(X_1-X_3)+\ldots \nonumber \\
w^2_3\,m_6\,h(m_6+1)               &=& (n_1 + n_2 - n_3 + 2\,m_6-1 )\,h()+2\,(X_2-X_1)+\ldots 
\label{h2solv}
\end{eqnarray}
where by $\ldots$ we denote the omitted boundary terms, which are proportional to $\delta_{n_i}$, and
\begin{eqnarray}
Q_1 &=& \frac{n_1\,(n_1-1)}{n_3+1}\,h(n_1-2,n_3+2)+\frac{n_1\,(n_1-1)}{n_2+1}\,h(n_1-2,n_2+2)+
\nonumber \\ && \frac{1}{n_3+1}\,h(n_3+2,m_5-1)+\frac{1}{n_2+1}\,h(n_2+2,m_6-1)\,,
\nonumber \\
Q_2 &=&\frac{n_2\,(n_2-1)}{n_3+1}\,h(n_2-2,n_3+2)+\frac{n_2\,(n_2-1)}{n_1+1}\,h(n_1+2,n_2-2)+
\nonumber \\ && \frac{1}{n_3+1}\,h(n_3+2,m_4-1)+\frac{1}{n_2+1}\,h(n_1+2,m_6-1)\,,
\nonumber \\
Q_3 &=&
\frac{n_3\,(n_3-1)}{n_2+1}\,h(n_2+2,n_3-2)+\frac{n_3\,(n_3-1)}{n_1+1}\,h(n_1+2,n_3-2)+
\nonumber \\ && \frac{1}{n_2+1}\,h(n_2+2,m_4-1)+\frac{1}{n_1+1}\,h(n_1+2,m_5-1)\,,
\label{Q}
\end{eqnarray}
\begin{eqnarray}
X_1 &=& (n_2-1)\,n_2\,m_4\,h(n_2-2,m_4+1)-(n_3-1)\,n_3\,m_4\,h(n_3-2,m_4+1)\,,\nonumber \\
X_2 &=& (n_3-1)\,n_3\,m_5\,h(n_3-2,m_5+1)-(n_1-1)\,n_1\,m_5\,h(n_1-2,m_5+1)\,,\nonumber \\
X_3 &=& (n_1-1)\,n_1\,m_6\,h(n_1-2,m_6+1)-(n_2-1)\,n_2\,m_6\,h(n_2-2,m_6+1)\,.
\label{X}
\end{eqnarray}
There is obviously some freedom in using Eqs. (\ref{heqs}) to simplify recursions,
and our choice allows to separate recursions into two subshells of
$n_1+n_2+n_3$ and $m_4+m_5+m_6$. We put $m_4=m_5=m_6=1$ in Eq. (\ref{h1solv}) and 
immediately obtain the first subset of recursions, Eq.~(\ref{rec1}). The second subset Eq. (\ref{rec2})
is obtained from Eq. (\ref{h2solv}) in three steps. The first step is 
setting $m_4=m_5=m_6=1$. The second step is multiplication of recursion formulas
by $w_i\,w_j$ to remove any $w_i$ from the denominator. The third step is
differentiation with respect to $w_1, w_2,$ and $w_3$, which converts 
the function $h()$ into the function $f()$. That completes the derivation of 
recursion relations for three-electron Hylleraas integrals.

\section{Summary}
We have derived a complete set of recursion relations for three-electron
Hylleraas integral $f$. They are sufficiently stable for a precise 
numerical calculation. 
Since all $w_i$ are bounded from below by
approximately $\sqrt{2\,E}$, where $E$ is the ionization energy,
the denominators of three-electron recursions
are also bounded from below. Therefore, by using
extended precision arithmetics (octuple precision) 
we can safely devote 1 digit per iteration, and still
preserve quad precision for about 30 iterations, which is the maximum
we aim to use. Possible instabilities in two-electron integrals are
avoided by applying inverse recursions \cite{kor1}. 
Numerical results are shown in Table \ref{table1}.
\begin{table}[!hbt]
\caption{Values of three-electron Hylleraas integral at $w_1=w_2=w_3=1$, 
$[m]$ denotes $10^m$}
\label{table1}
\begin{ruledtabular}
\begin{tabular}{rll}
      $n$ & $f(n,0,0;0,0,0)$           &  $f(0,0,0;n,0,0)$  \\
\hline
      0   &   2.208\,310\,154\,388\,618\,874\,536\,424[-1] &      2.208\,310\,154\,388\,618\,874\,536\,424[-1] \\
      1   &   2.876\,820\,724\,517\,809\,274\,392\,190[-1] &      2.208\,310\,154\,388\,618\,874\,536\,424[-1] \\
      2   &   6.071\,253\,765\,587\,525\,062\,881\,067[-1] &      3.658\,582\,716\,243\,175\,207\,969\,277[-1] \\
      3   &   1.801\,456\,579\,614\,247\,419\,513\,752[0] &       8.803\,723\,087\,040\,150\,596\,505\,449[-1] \\
      4   &   6.949\,688\,537\,201\,117\,333\,162\,822[0] &       2.849\,464\,173\,126\,685\,211\,199\,798[0] \\
      5   &   3.316\,893\,553\,498\,521\,645\,367\,878[1] &       1.176\,795\,411\,671\,425\,935\,279\,582[1] \\
      6   &   1.892\,427\,697\,247\,010\,803\,401\,964[2] &       5.962\,899\,567\,501\,152\,778\,486\,008[1] \\
      7   &   1.258\,719\,915\,821\,483\,876\,136\,660[3] &       3.596\,955\,116\,745\,326\,378\,301\,909[2] \\
      8   &   9.575\,385\,319\,725\,442\,534\,735\,866[3] &       2.522\,862\,411\,307\,814\,783\,043\,058[3] \\
      9   &   8.206\,804\,555\,680\,135\,296\,239\,238[4] &       2.019\,476\,554\,953\,447\,619\,512\,494[4] 
\end{tabular}
\end{ruledtabular}
\end{table}
These recursion relations can be used for high precision calculation
of lithium wave function with a small computational effort. 
We think that similar recursions can be derived also for beryllium, which 
involves four
electrons, but the generic integral with all powers of $r_{ij}$ equal to $-1$
has not yet been obtained. However, this generic integral can be represented
as a multiple sum as recently indicated by Sims and Hagstrom \cite{sims} and Frolov 
\cite{frolov} and thus obtained numerically. Coming back to lithium, our aim is 
the high precision calculation of higher order relativistic and QED effects
like that for lithium hyperfine splitting. It requires, however, also the 
calculation
of Hylleraas integrals with various negative powers of $r_i$ and
$r_{ij}$. $1/r_i^n$ with $n>1$ can be expressed in terms of harmonic polylogarithms,
which were introduced recently in~\cite{har}, 
and $1/r_{ij}^n$ can be obtained, we think, 
by generalization of the recursion relations obtained in
this work. \par 
In summary, we have presented a simple method to derive recursion
relations and solved them for three-electron Hylleraas integral.

\section{Acknowledgments}
We are grateful to Vladimir Korobov for his source code
of the fast multiprecision arithmetics and of two-electron 
recursions in the singular case. This work was supported by EU grant
HPRI-CT-2001-50034.

\appendix
\renewcommand{\theequation}{\Alph{section}\arabic{equation}}
\renewcommand{\thesection}{\Alph{section}}
\section{Definitions of functions and relations}
\setcounter{equation}{0}
We use the symbol $\delta_n$ which denotes the Kronecker $\delta_{n,0}$. 
Definitions of all functions used in this work are presented below.
$r_{ij}$ denotes $|\vec r_i-\vec r_j|$ and $k_{ij}=|\vec{k_i}-\vec{k_j}|$. 
All the $n_i$ and $m_i$ are assumed to be
nonnegative integer and $w_i, u_i$ to be real and positive. 
\begin{eqnarray}
F(n_1,n_2,n_3;n_4,n_5,n_6) &=& \int\frac{d^3\,r_1}{4\,\pi}\int\frac{d^3\,r_2}{4\,\pi}
\int\frac{d^3\,r_3}{4\,\pi}\,e^{-w_1\,r_1-w_2\,r_2-w_3\,r_3-u_1\,r_{23}-u_2\,r_{31}-u_3\,r_{12}}
\nonumber \\ &&
r_{23}^{n_1-1}\,r_{31}^{n_2-1}\,r_{12}^{n_3-1}\,
r_{1}^{n_4-1}\,r_{2}^{n_5-1}\,r_{3}^{n_6-1} \\\nonumber \\
f(n_1,n_2,n_3;n_4,n_5,n_6) &=& \int\frac{d^3\,r_1}{4\,\pi}\int\frac{d^3\,r_2}{4\,\pi}
\int\frac{d^3\,r_3}{4\,\pi}\,e^{-w_1\,r_1-w_2\,r_2-w_3\,r_3} \nonumber \\ &&
r_{23}^{n_1-1}\,r_{31}^{n_2-1}\,r_{12}^{n_3-1}\,
r_{1}^{n_4-1}\,r_{2}^{n_5-1}\,r_{3}^{n_6-1}\\ \nonumber \\
f(\star,n_2,n_3;n_4,n_5,n_6) &=& \int\frac{d^3\,r_1}{4\,\pi}\int\frac{d^3\,r_2}{4\,\pi}
\int\frac{d^3\,r_3}{4\,\pi}\,e^{-w_1\,r_1-w_2\,r_2-w_3\,r_3}\nonumber \\ &&
4\,\pi\,\delta^3(r_{23})\,r_{31}^{n_2-1}\,r_{12}^{n_3-1}\,
r_{1}^{n_4-1}\,r_{2}^{n_5-1}\,r_{3}^{n_6-1}\\\nonumber \\
f(n_1,n_2,n_3;\star,n_5,n_6) &=&  \int\frac{d^3\,r_1}{4\,\pi}\int\frac{d^3\,r_2}{4\,\pi}
\int\frac{d^3\,r_3}{4\,\pi}\,e^{-w_1\,r_1-w_2\,r_2-w_3\,r_3} \nonumber \\ &&
r_{23}^{n_1-1}\,r_{31}^{n_2-1}\,r_{12}^{n_3-1}\,
4\,\pi\,\delta^3(r_1)\,r_{2}^{n_5-1}\,r_{3}^{n_6-1}\\\nonumber \\
\Gamma(n_1,n_2,n_3;\alpha_1,\alpha_2,\alpha_3) &=&  \int\frac{d^3\,r_1}{4\,\pi}
\int\frac{d^3\,r_2}{4\,\pi}\,e^{-\alpha_1\,r_1-\alpha_2\,r_2-\alpha_3\,r_{12}}\,
r_{1}^{n_1-1}\,r_{2}^{n_2-1}\,r_{12}^{n_3-1}\label{A5}\\\nonumber \\
G(m_1,m_2,m_3;m_4,m_5,m_6) &=& \frac{1}{8\,\pi^6}\,\int d^3k_1\int d^3k_2\int d^3k_3\,
(k_1^2+u_1^2)^{-m_1}\,(k_2^2+u_2^2)^{-m_2} \label{A6}\\ &&(k_3^2+u_3^2)^{-m_3}\,
(k_{32}^2+w_1^2)^{-m_4}\,(k_{13}^2+w_2^2)^{-m_5}\,(k_{21}^2+w_3^2)^{-m_6}
\nonumber \\\nonumber \\ 
h(n_1,n_2,n_3;m_4,m_5,m_6) &=& 
(-1)^{n_1}\,\frac{\partial^{n_1}}{\partial u_1^{n_1}}\biggr|_{u_1=0}\,
(-1)^{n_2}\,\frac{\partial^{n_2}}{\partial u_2^{n_2}}\biggr|_{u_2=0}\,
(-1)^{n_3}\,\frac{\partial^{n_3}}{\partial u_3^{n_3}}\biggr|_{u_3=0}\nonumber \\ &&
G(1,1,1,m_4,m_5,m_6) \label{A7}
\\\nonumber \\
f(n_1,n_2,n_3;n_4,n_5,n_6) &=& 
(-1)^{n_4}\,\frac{\partial^{n_4}}{\partial w_1^{n_4}}\,
(-1)^{n_5}\,\frac{\partial^{n_5}}{\partial w_2^{n_5}}\,
(-1)^{n_6}\,\frac{\partial^{n_6}}{\partial w_3^{n_6}}\,h(n_1,n_2,n_3,1,1,1)\nonumber \\
\label{def}
\end{eqnarray}

\end{document}